\documentclass[12pt]{article}

\usepackage{amsmath,amsthm, amssymb}

\newcommand{\be}{\begin{equation}}
\newcommand{\ee}{\end{equation}}
\newcommand{\bea}{\begin{eqnarray}}
\newcommand{\eea}{\end{eqnarray}}
\newcommand{\nn}{\nonumber}

\newcommand{\pr}{\prime}

\newcommand{\wt}{\widetilde}
\newcommand{\Psitil}{\wt \Psi}

\newcommand{\Eta}{\mathcal H}
\newcommand{\Etatil}{\wt \Eta}

\newcommand{\hf}{{\hat5}}
\newcommand{\p}{\partial}

\newcommand{\al}{\alpha}
\newcommand{\la}{\lambda}
\newcommand{\da}{\delta}
\newcommand{\om}{\omega}
\newcommand{\Ga}{\Gamma}
\newcommand{\Si}{\Sigma}
\newcommand{\si}{\sigma}

\title{
{\bf Boundary Conditions in Brane\,-World Supergravity}\thanks{
Talk given at International Workshop ``Supersymmetries and Quantum
Symmetries'' (Dedicated to the 75th anniversary of the birth of
Victor Isaakovich Ogievetsky); JINR, Dubna; July 24-29, 2003}
}
\author{
 \large{\bf Jonathan A.~Bagger}~\thanks{bagger@jhu.edu} 
 \quad \it{and} \quad    
 \large{\bf Dmitry V.~Belyaev}~\thanks{belyaev@pha.jhu.edu}\\[5mm]
 \it Department of Physics and Astronomy,\\ 
 \it The Johns Hopkins University,\\
 \it 3400 North Charles Street,\\
 \it Baltimore, MD 21218, USA}
\date{}

\begin{document}

\numberwithin{equation}{section}

\maketitle

\begin{abstract}
\noindent
Bulk supergravity on a manifold with boundary must be supplemented by boundary conditions that preserve local supersymmetry.  This ``downstairs'' picture has certain advantages over the equivalent ``upstairs'' picture, expressed in terms of orbifolds.  In particular, Scherk-Schwarz supersymmetry breaking can be described much more simply in the downstairs picture.  Nevertheless, physics on the fundamental domain can always be lifted upstairs, so long as fields are allowed to be discontinuous across the boundary.  In this talk we apply these considerations to five-dimensional supergravity in a warped Randall-Sundrum background.
\end{abstract}



\section{Motivation and Results}

Brane-world scenarios of the Randall-Sundrum type provide a geometrical solution to the gauge hierarchy problem \cite{rs}.  The scenarios are rooted in string theory, so it is natural to expect that they can be made supersymmetric.  Indeed, supersymmetric versions of the Randall-Sundrum scenario were constructed by three groups \cite{abn, gp, flp}.  The essential difference is that in Refs.~\cite{gp, flp}, all fields are assumed to be continuous across the branes, while in Ref.~\cite{abn}, the fermionic fields are allowed to jump.  

In Ref.~\cite{bb1}, we carried out a more general supersymmetrization of the Randall-Sundrum setup, parametrized by a unit vector, $\vec q$.  Our approach expands the space of allowed supersymmetrizations.  We recover the results of Refs.~\cite{abn} and \cite{gp, flp} as special cases, corresponding to $\vec q =(-1,0,0)$ and $\vec q=(0,0,1)$, respectively.   In addition, our approach allows us to supersymmetrize the ``detuned'' Randall-Sundrum scenario, proving that the Randall-Sundrum fine-tuning is not a consequence of supersymmetry.

Our results are easily derived in the ``downstairs'' picture, where all fields are defined on the fundamental interval, instead of in the ``upstairs'' picture, where the fields are lifted to an orbifold.  In the downstairs picture, all the fields are continuous.  From this point of view, the constructions of Refs.~\cite{abn, gp, flp} differ only in the boundary conditions imposed at the ends of the interval.

In Ref.~\cite{bb2}, we extended this work by studying the question of spontaneous supersymmetry breaking via the Scherk-Schwarz mechanism.  As before, we found the analysis to be easiest in the downstairs picture.  We identified the locally supersymmetric boundary conditions for which global supersymmetry is broken.  We demonstrated that supersymmetry can be spontaneously broken by boundary conditions in the detuned scenario, but not in the original case with tuned brane tensions.

We also showed explicitly how to lift to the upstairs picture, where the conventional description of the Scherk-Schwarz mechanism applies.  This lifting can be quite complicated, with jumping and/or non-periodic fields.  All the complications can be avoided by working in the downstairs picture.  The physics is determined by the theory (including boundary conditions) on the fundamental interval, irrespective of how the theory is lifted to the covering space.

\section{Randall-Sundrum Model}

For the purpose of this talk, the Randall-Sundrum scenario~\cite{rs} is a theory of pure gravity on a five-dimensional manifold with two parallel branes.  The action is
\be
S=\int d^5x\ e_5\big(-\frac{1}{2}R-\Lambda_5\big)
-\int_{\Sigma_1} d^4x\ e_4 T_1
-\int_{\Sigma_2} d^4x\ e_4 T_2,
\ee
where $\Lambda_5$ is the bulk cosmological constant and $T_{1,2}$ are the tensions of the 3-branes $\Sigma_{1,2}$. With the parametrization
\be
\Lambda_5=-6\la^2, \quad 
T_1=6\la_1, \quad
T_2=-6\la_2,
\ee
the condition for the tuned Randall-Sundrum scenario is
\be
\la_1=\la_2=\la.
\ee
The Randall-Sundrum background has flat four-dimensional slices,
\be
ds^2=a^2(z)\eta_{mn}dx^m dx^n+dz^2,
\ee
corresponding to an effective theory with zero cosmological constant, $\Lambda_4=0$. In these coordinates, the warp factor $a(z)$ is an exponential,
\be
a(z)=\exp(-\la |z|),
\ee
which gives rise to an exponential hierarchy between the fundamental energy scales on the two branes.

In the downstairs picture, the fifth coordinate is taken to be in the range $z\in [0,\pi R]$, and the two branes sit at the ends of the interval.  In the upstairs picture, $z\in [-\pi R,\pi R]$, where the two ends are identified and the fifth dimension has topology of a circle.  The doubling of the fundamental interval is a mathematical trick.  In the orbifold construction, the fields on one half of the circle determine the fields on the other half.  This requires them to have a definite parity under the reflection $z \rightarrow -z$.

\section{Bulk Supergravity}

The supersymmetrization of pure gravity in the five-dimensional bulk is a gauged supergravity, where in addition to graviton, $e_M^A$, there is a graviphoton, $B_M$, and a gravitino, $\Psi_M^i$. The index $i$ on the spinors is a two-dimensional index for $SU(2)$ $R$-symmetry, a $U(1)$ subgroup of which, parametrized by $\vec q$, is gauged. The action is 
\bea
&&S_{\rm bulk} = \int d^5x\ e_5
\Big\{
-\frac{1}{2} R -\frac{1}{4}F_{MN}F^{MN}
+\frac{i}{2}\Psitil_M^i \Ga^{MNK} D_N \Psi_{Ki}
\nn\\
&&\qquad\qquad\qquad\qquad
+ 6\la^2
-\frac{3}{2} \la\, \vec q\cdot \vec\si_i{}^j \Psitil_M^i\Si^{MN}\Psi_{Nj}
+\dots
\Big\}.
\eea
(See Ref.~\cite{bb1} for details.)  It is invariant under the following local $N=2$ supersymmetry
\bea
\da e_M^A &=& i \Etatil^i\Ga^A\Psi_{Mi}, \quad
\da B_M = i\frac{\sqrt6}{2}\Etatil^i\Psi_{Mi} \nn\\
\da \Psi_{Mi} &=& 2 \big(
D_M\Eta_i - i\frac{\sqrt6}{2}\la\, \vec q\cdot \vec\si_i{}^j B_M\,\Eta_j 
\big) +\dots
\eea
The supersymmetry parameter $\Eta^i$ is a symplectic-Majorana spinor, so it contains two two-component spinors $\eta_1$ and $\eta_2$.  The action is invariant (up to a boundary term) when $\eta_{1,2}(x,z)$ are arbitrary functions of the five-dimensional coordinates. 

To lift the theory to the orbifold, we need to specify a $\mathbb{Z}_2$ symmetry.  For each field $\phi$, we must identify a parity $P(\phi)=\pm 1$, so that
\be
z^\pr=-z, \quad
\phi^\pr(x,z)=P(\phi)\phi(x,z)
\ee
is a symmetry of the action.  We choose the following parity assignments:
\bea
&&P(e_m^a, e_5^\hf, B_5, \eta_1, \psi_{m1}, \psi_{52};
q_{1,2}, \la)=+1 \\
&&P(e_m^\hf, e_5^a, B_m, \eta_2, \psi_{m2}, \psi_{51};
q_3)=-1.
\eea
The fact that $q_3$ has negative parity means that the $\mathbb{Z}_2$ symmetry is, in fact, broken for $q_3\neq 0$.  The symmetry can be ``restored'', however, if we allow $q_3$ to flip as well.  This is what gives rise to the ``odd bulk mass term'' in Refs.~\cite{gp, flp}.

When the bulk cosmological constant, $\Lambda_5=-6\la^2$, is zero, the action is invariant under an $SU(2)$ $R$-symmetry,
\be
\Psi_{Mi}^\pr=U_i^j\Psi_{Mj},
\ee
where $U\in SU(2)$.  When the cosmological constant is not zero, the $SU(2)$ is broken to $U(1)$, parametrized by $\vec q$.  The residual invariance is given by $U=\exp(i\vec q\cdot\vec\si)$.  The full $SU(2)$ symmetry can, however, be ``restored'' if we rotate the parameters $\vec q=(q_1, q_2, q_3)$ as follows,
\be
\Psi_M^\pr=U\Psi_M, \quad
(\vec q^{\,\pr}\cdot\vec\si)=U(\vec q\cdot\vec\si)U^\dagger.
\ee
This is completely analogous to flipping $q_3^\pr=-q_3$ in the case of the $\mathbb{Z}_2$ symmetry.  In Ref.~\cite{bb2} we showed how the ``restored'' $SU(2)$ symmetry can be used to lift from the downstairs to the upstairs picture.

\section{Boundary Conditions}

Under local supersymmetry, the supergravity Lagrangian varies into a total derivative. Therefore, on a manifold with boundary, the action varies into a boundary term,
\be
\da S_{bulk}=\int_{\mathcal{M}}\p_M K^M
=\int_{\p\mathcal{M}}(n_M K^M)
=\int_{z=\pi R}K^5(x, \pi R) -\int_{z=0}K^5(x,0).
\ee
To make the boundary variation vanish, we need to specify boundary conditions for the fields. Such boundary conditions, sufficient for warped backgrounds, are given by
\bea
&&e_5^a=e_m^\hf=B_m=0, \quad
\om_{ma\hf}=\la_i e_{ma}\\
&&\eta_2=\al_i\eta_1, \quad
\psi_{m2}=\al_i\psi_{m1}, \quad
\psi_{51}=-\al_i^\ast\psi_{52}
\eea
on $\Si_i$, provided the parameters $\la_{1,2}\in\mathbb{R}$ and $\al_{1,2}\in\mathbb{C}$ satisfy
\be
\label{laqal}
\la_i=-\frac{(\al_i+\al_i^\ast)q_1+i(\al_i^\ast-\al_i)q_2
+(\al_i\al_i^\ast-1)q_3}{1+\al_i\al_i^\ast}\la.
\ee
One can check that these conditions are consistent with supersymmetry and that they lead to $K^5(x,z)=0$ for $z=0, \pi R$.  These conditions are all one needs for supersymmetry invariance in the downstairs picture.

The bosonic boundary conditions correspond, in the upstairs picture, to setting the odd bosonic fields to zero on the branes.  The fermionic boundary conditions are dictated by supersymmetry.  They allow only one free parameter $\al_i$ for each brane.  When  $\al_i \ne 0$, the odd fermionic fields jump on the $i^{\rm th}$ brane.  Note that the restriction (\ref{laqal}) implies that $\la_{1,2}$ are bounded by $|\la_i|\leq \la$.

In the upstairs picture, the boundary conditions and parity assignments determine the jumps (or discontinuities) for the fields.  These jumps follow from a brane action by matching delta-function singularities in the equations of motion.  In our case, the brane action is given by
\bea
S_{brane}=
\int_{\Si_1}d^4x\ e_4\big[
-6\la_1-(2\al_1\psi_{m1}\si^{mn}\psi_{m1}+\text{h.c.})\big]\nn\\
-\int_{\Si_2}d^4x\ e_4\big[
-6\la_2-(2\al_2\psi_{m1}\si^{mn}\psi_{m1}+\text{h.c.})\big].
\eea
Therefore, in the upstairs picture, we see that $T_1=6\la_1$ and $T_2=-6\la_2$ are the brane tensions, while $\al_{1,2}$ are brane-localized gravitino mass terms.

Checking supersymmetry in the upstairs picture is tricky since one encounters discontinuous fields and products of distributions.  One can see how this works in Ref.~\cite{bb1}.   But if there are any doubts, one can always go back to the downstairs picture and check everything there.  The two pictures are equivalent, so a proof in either picture should suffice.

\section{Supersymmetry Breaking}

The tuned case, with $\la_1=\la_2=\la$, gives rise to a bosonic background with flat ($\Lambda_4=0$) four-dimensional slices.  The detuned case, with $|\la_{1,2}|<\la$, give rise to backgrounds with anti de-Sitter slices ($\Lambda_4<0$).  For each case one can ask whether supersymmetry is preserved by the background, or whether it is spontaneously broken.

Supersymmetry is preserved if the Killing spinor equations have a solution.  On the manifold with boundary, the solution must obey appropriate boundary conditions, which in our case are parametrized by $\al_{1,2}$.  In Ref.~\cite{bb2} we demonstrated that Killing spinors always exist for the tuned case; the $\al_{1,2}$ are fixed by the brane tensions and supersymmetry is not broken.\footnote{Compare this result to the case with no warping ($\la_1=\la_2=\la=0$), where brane-localized gravitino mass terms can be used to spontaneously break supersymmetry \cite{bfz}.}  For the detuned case, however, there is a one-parameter family of boundary conditions that break supersymmetry.  This relies on the fact that the $\al_{1,2}$ can be complex.

In Ref.~\cite{bb2} we lifted this setup to the upstairs picture, where it corresponds to supersymmetry breaking by the Scherk-Schwarz mechanism.  In fact, the lifting is not unique.  The general lifting uses a twist that corresponds to an element of the broken part of $SU(2)$.  The $SU(2)$ is restored by an appropriate choice of parameters $\vec q$ on different parts of $S^1$ or $\mathbb{R}$.

This work was supported in part by the U.S. National Science Foundation, grant NSF-PHY-9970781.

\end{document}